\documentclass[conference]{IEEEtran}
\IEEEoverridecommandlockouts
\usepackage{cite}
\usepackage{amsmath,amssymb,amsfonts}
\usepackage{algorithmic}
\usepackage{graphicx}
\usepackage{textcomp}
\usepackage{xcolor}
\def\BibTeX{{\rm B\kern-.05em{\sc i\kern-.025em b}\kern-.08em
    T\kern-.1667em\lower.7ex\hbox{E}\kern-.125emX}}
\usepackage{multirow}
\usepackage{array}
\usepackage{booktabs}
\usepackage{xcolor}
\usepackage{flushend}
\usepackage{setspace}
\begin{document}

\title{A New OTFS-Based Index Modulation System for 6G and Beyond: OTFS-Based Code Index Modulation  
\thanks{This work was supported by TUBITAK 1001 (Grant Number: 123E513).}}

\author{Author Name}

\author{\IEEEauthorblockN{Burak Ahmet Ozden\IEEEauthorrefmark{1}\IEEEauthorrefmark{2},
Erdogan Aydin\IEEEauthorrefmark{1}, 
Emir Aslandogan\IEEEauthorrefmark{3}, Haci Ilhan\IEEEauthorrefmark{3}, \\
Ertugrul Basar\IEEEauthorrefmark{4}\IEEEauthorrefmark{5}, 
Miaowen Wen\IEEEauthorrefmark{6}, 
Marco Di Renzo\IEEEauthorrefmark{7}}
\IEEEauthorblockA{\IEEEauthorrefmark{1}Department of Electrical and Electronics Engineering, Istanbul Medeniyet University, Istanbul, Turkey.}
\IEEEauthorblockA{\IEEEauthorrefmark{2}Department of Computer Engineering, Yildiz Technical University, Istanbul, Turkey.}
\IEEEauthorblockA{\IEEEauthorrefmark{3}Department of Electronics and Communication Engineering, Yildiz Technical University, Istanbul, Turkey.}
\IEEEauthorblockA{\IEEEauthorrefmark{4}Department of Electrical Engineering, Tampere University, 33720 Tampere, Finland.} 
\IEEEauthorblockA{\IEEEauthorrefmark{5}Department of Electrical and Electronics Engineering, Koc University, Istanbul, Turkey.} 
\IEEEauthorblockA{\IEEEauthorrefmark{6}School of Electronic and Information Engineering, South China University of Technology, Guangzhou, China.}
\IEEEauthorblockA{\IEEEauthorrefmark{7} Laboratoire des Signaux et Syst\`emes, CentraleSup\'elec, CNRS, Universit\'e Paris-Saclay, France.}
\\  Email: bozden@yildiz.edu.tr, erdogan.aydin@medeniyet.edu.tr, emira@yildiz.edu.tr,  ilhanh@yildiz.edu.tr,  \\
ebasar@ku.edu.tr, eemwwen@scut.edu.cn, marco.di-renzo@universite-paris-saclay.fr.
}

\maketitle

\begin{abstract}

This paper proposes the orthogonal time frequency space-based code index modulation (OTFS-CIM) scheme, a novel wireless communication system that combines OTFS modulation, which enhances error performance in high-mobility Rayleigh channels, with CIM technique, which improves spectral and energy efficiency, within a single-input multiple-output (SIMO) architecture. The proposed system is evaluated through Monte Carlo simulations for various system parameters. Results show that increasing the modulation order degrades performance, while more receive antennas enhance it. Comparative analyses of error performance, throughput, spectral efficiency, and energy saving demonstrate that OTFS-CIM outperforms traditional OTFS and OTFS-based spatial modulation (OTFS-SM) systems. Also, the proposed OTFS-CIM system outperforms benchmark systems in many performance metrics under high-mobility scenarios, making it a strong candidate for sixth generation (6G) and beyond.

\end{abstract} 


\vspace{-0.5 em}
\section{Introduction}

\begin{table*}[t]
\centering
\addtolength{\tabcolsep}{-4pt}
\caption{Mapping procedure for the proposed OTFS-CIM system.}\vspace{-0.5em}
 
\label{OTFS_CIM_MAP}
\begin{tabular}{|c|c|c|c|c|c|} 
\hline
\hline
\textbf{\small Parameters}       & \textbf{\begin{tabular}[c]{@{}c@{}}\small Data Bits\\ \end{tabular}} & \textbf{\begin{tabular}[c]{@{}c@{}} \small First DD grid \\ \end{tabular}} & \textbf{\begin{tabular}[c]{@{}c@{}} \small Second DD grid \\ \end{tabular}} & \textbf{\begin{tabular}[c]{@{}c@{}} \small ... \\ \end{tabular}} & \textbf{\begin{tabular}[c]{@{}c@{}} \small Sixteenth DD grid \\ \end{tabular}} \\ \hline\hline
\begin{tabular}[c]{@{}c@{}} \small \!$M_q=4$,\! $N_{C}=2$,\!  \\ \!$N=4$,\! $M=4$ \!\end{tabular} & 
\begin{tabular}[c]{@{}c@{}} {[}00011001...1111{]} \\ $n_{OTFS-CIM}=64$ \end{tabular}                                                             & \begin{tabular}[c]{@{}c@{}}{[}0001{]}\\  \!$s_\Re[1,1] =-1$, \!$s_\Im[1,1] =-1$ \! \\ $c_\Re=1$, $c_\Im=1$ \end{tabular}                   & \begin{tabular}[c]{@{}c@{}}{[}1001{]} \\  \!$s_\Re[1,2] =-1$,\! $s_\Im[1,2] =-1$\!  \\ $c_\Re=2$, $c_\Im=1$ \end{tabular}                 & \begin{tabular}[c]{@{}c@{}}...\\ ... \\ ...  \end{tabular}  & \begin{tabular}[c]{@{}c@{}}{[}1111{]} \\ \!\!\! $s_\Re[4,4] =1$, \!$s_\Im[4,4] =-1$ \!\!\! \\ $c_\Re=2$, $c_\Im=2$\! \end{tabular} \\ \hline \hline
\end{tabular}
\vspace{-1.5em}
\end{table*}

The rapid increase in internet-connected devices, mobile users, autonomous vehicles, and intelligent systems has led to the emergence of new and challenging requirements in wireless communications. Among these requirements, high data rate, low latency, high capacity, high spectral and energy efficiency, reliable communication, and low error rate data transmission are the most important features. High-resolution video streams, the sharing of large data, and virtual reality applications have increased the demand for fast data transmission. The rapid increase in the number of Internet of Things (IoT) applications and mobile users has increased the need for larger data capacity, while the long-lasting battery usage and environmentally friendly designs of devices have made energy efficiency an important requirement \cite{6G1,RISantenna,6G3}. Existing wireless communication systems are insufficient to meet all these requirements and realize next-generation networks, such as the sixth generation (6G). Therefore, developing innovative and high-performance wireless communication systems has become mandatory.

Index modulation (IM), a novel and efficient approach for wireless data transmission, has garnered significant attention due to its notable advantages. IM enables the transmission of additional information by embedding data in the symbols and the indices of system parameters used in wireless communication. These parameters include antennas, subcarriers, time slots, radio frequency (RF) mirrors, relays, spreading codes, etc. IM significantly improves spectral and energy efficiency by utilizing the indices of such parameters as part of the modulation space. Since the data mapped to the indices is embedded within the transmitted signal, minimal energy is required for the extra data transmission. Moreover, IM systems offer key advantages, including the elimination of inter-channel interference (ICI), reduced system complexity, improved error performance, and minimized energy consumption \cite{IM1, IM2, IM3}.

Code index modulation (CIM) is a significant IM technique that utilizes the indices of orthogonal spreading codes to enable faster and more reliable data transmission with reduced errors. The CIM technique enhances data security and resilience against noise while leveraging conventional modulation techniques to achieve higher data rates. It is widely employed in various wireless communication systems, including cellular networks, satellite communications, and wireless local area networks. Compared to traditional techniques, the CIM approach facilitates more efficient data transmission, resulting in more reliable, faster communication with lower energy consumption and improved performance \cite{CIM1, CIM3}. The study of \cite{CIM4} introduces the generalized CIM (GCIM) technique, which employs spreading codes for data transmission in high-rate wireless communication systems. The proposed GCIM method outperforms traditional modulation techniques and the spatial modulation (SM) technique regarding error performance. Additionally, the study of \cite{CIM6} introduces an innovative communication system called CIM-SMBM, which combines SM, media-based modulation (MBM), and CIM into a multiple-input multiple-output (MIMO) system. The proposed system demonstrates improved energy efficiency, enhanced spectral efficiency, better error performance, and higher data rates compared to benchmark systems.

Orthogonal frequency division multiplexing (OFDM) is widely utilized in current technologies due to its ability to mitigate inter-symbol interference (ISI) caused by channel delays, but it faces limitations in high-mobility scenarios. In such environments, adverse effects like multipath fading, ISI, delays, and Doppler shifts result in the loss of orthogonality among subcarriers, leading to ICI and degraded performance. By operating in the delay-Doppler (DD) domain, OTFS provides robust communication, offering significant advantages in handling time-varying channels and high performance in next-generation wireless systems \cite{SURVEY1}. In recent studies, IM-based orthogonal time frequency space (OTFS) models significantly enhance system performance across multiple metrics. In \cite{OTFS_IM1}, the proposed joint DD index modulation-based OTFS (JDDIM-OTFS) scheme activates DD elements with distinct constellations and employs a greedy detector, achieving better BER performance than conventional OTFS. The study of \cite{OTFS_IM2} proposes two new block-wise IM schemes for OTFS: Delay-IM with OTFS (DeIM-OTFS) and Doppler-IM with OTFS (DoIM-OTFS), which enable block-wise delay/Doppler resource activation, with maximum likelihood (ML) and low-complexity detection algorithms showing robustness to imperfect CSI. Similarly, the study of \cite{OTFS_IM3} focuses on DoIM-OTFS, activating Doppler bins using a customized message passing (CMP) algorithm for superior BER performance under imperfect CSI. Also, recent studies have introduced various spatial modulation-based OTFS (SM-OTFS) systems, including an SM-OTFS scheme for improved spectral efficiency and lower detection complexity in \cite{OTFS_SM1}, an SM-OTFS system designed for high transmission reliability with theoretical error rate analysis in \cite{OTFS_SM2}, and an SM-OTFS scheme utilizing a low-complexity distance-based detection algorithm in \cite{OTFS_SM3}.

This paper introduces a new wireless communication system called OTFS-CIM, which combines the OTFS technique, which improves error performance in high mobility channel situations, and the CIM technique, which provides high spectral and energy efficiency, in a single-input multiple-output (SIMO) architecture. The contributions of the paper can be expressed as follows:
\begin{enumerate}

\item The simulation results of the proposed CIM-OTFS system are obtained using the Monte Carlo technique over high-mobility Rayleigh channels for $M_q$-quadrature amplitude modulation (QAM), considering various system parameters.

\item The effects of the modulation order and the number of receive antennas on the performance of the proposed CIM-OTFS system are investigated. It is observed that increasing the modulation order degrades system performance, while increasing the number of receiver antennas enhances it.

\item The throughput, spectral efficiency, and energy-saving analyses of the proposed OTFS-CIM system are conducted and compared with OTFS and OTFS-SM systems. It is observed that the OTFS-CIM system outperforms the other systems in all analyses.

\item The error performance of the proposed OTFS-CIM system is presented under various conditions and compared with OTFS and OTFS-SM systems. The results demonstrate that the proposed system achieves superior error performance in all scenarios.

\end{enumerate}

The rest of this paper is as follows: Section II introduces the proposed OTFS-CIM system model. Section III provides performance analyses, including throughput, spectral efficiency, and energy-saving analyses. Section IV presents simulation results, and Section V concludes the paper.

The notations used in this paper are as follows. Vectors/matrices are denoted by bold lowercase/uppercase characters. Also, $(\cdot)^T$, $\|\cdot\|$, and $\otimes$ represent the transpose, Euclidean norm, and Kronecker product, respectively.

\begin{figure}[t]
\centering{\includegraphics[width=0.5\textwidth]{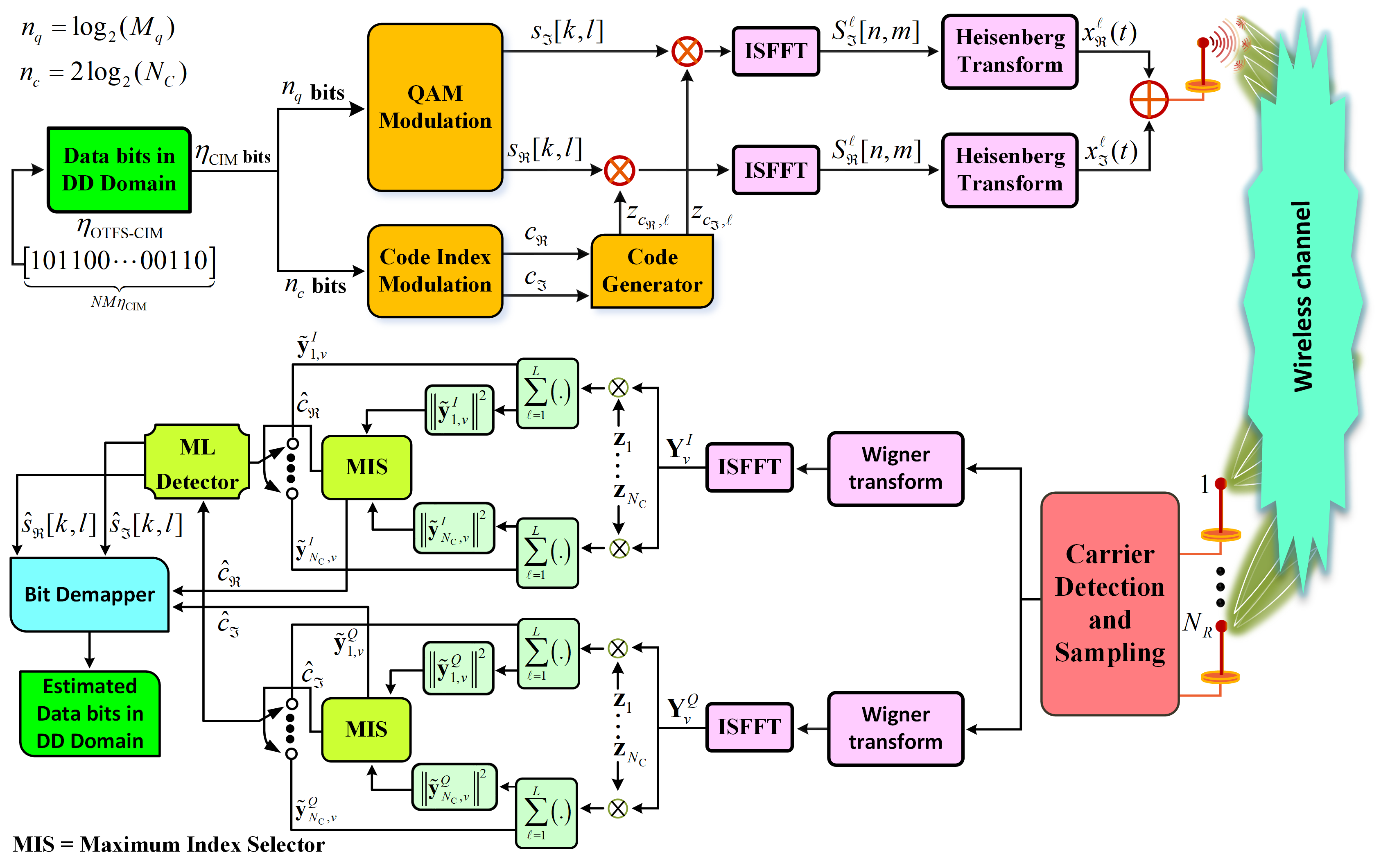}}	\vspace{-1em}
	\caption{System model of the proposed OTFS-CIM system.}
	\label{system_model} 
    \vspace{-1.5em}
\end{figure}

\vspace{-0.5em}
 
\section{The proposed OTFS-CIM System Model}

Fig. \ref{system_model} shows the overall system model structure of the proposed OTFS-CIM scheme. The proposed system model has one transmit antenna at the transmitter and $N_R$ receive antenna at the receiver. Furthermore, the proposed OTFS-CIM system uses $N_C$ spreading codes, which are orthogonal Walsh-Hadamard (WH) codes $\textbf{z}_{c} = [z_{c, 1}, z_{c, 2}, \ldots, z_{c, L}]^T$, where $c \in\{1,2, ..., N_C\}$, and each code consists of $L$ chips. That is, $N_C$ spreading codes of length $L$ are used. Also, the  $\ell^\text{th}$ chip of the $c^\text{th}$ spreading code $z_{c,\ell} \in \{\frac{1}{\sqrt{L}},\frac{-1}{\sqrt{L}}\}$ and $\ell \in \{1,2, ..., L\}$. For each of the in-phase ($I$) and quadrature ($Q$) components, two spreading codes ($\textbf{z}_{c_\Re}$, $\textbf{z}_{c_\Im} \in \mathbb{R}^{L \times 1} $) are selected from the $N_C$ spreading code according to the data bits to be transmitted at the spreading code indices in each transmission period. Furthermore, in each transmission period, a symbol ($s=s_\Re+j s_\Im$) is selected from the $M_q$-QAM symbols according to the data bits to be carried in the symbols, where $M_q$ is the modulation order. Therefore, the spectral efficiency, which is the number of bits transmitted in a traditional CIM system during a transmission period, is expressed as follows:
\begin{equation}
    \eta_\text{{CIM}}=\log_2(M_q) + 2\log_2(N_C).
\end{equation}
Since a CIM-OTFS transmission frame consists of $N$ independent $M_q$-QAM symbols transmitted on $M$ subcarriers and the real and imaginary parts of the symbols are multiplied by $\textbf{z}_{c_\Re}$ and $\textbf{z}_{c_\Im}$ spreading codes chosen from the $N_C$ spreading codes, the spectral efficiency of the proposed CIM-OTFS system in bits per channel use (bpcu) for a DD grid of size ($N \times M$) can be defined as follows:
\begin{equation}
    \eta_\text{{OTFS-CIM}}=NM\Big(\log_2(M_q) + 2\log_2(N_C) \Big).
    \label{spec_cim}
\end{equation}

In the OTFS-CIM system, the \( MN \) information symbols are represented as \( s[k, l] \), where \( k = 0, \ldots, N-1 \) and \( l = 0, \ldots, M-1 \), and are arranged over the DD grid. Here, \( M \) denotes the
number of subcarriers, while \( N \) represents the time slots. The subcarrier spacing is \( \Delta f = \frac{1}{T_c} \). The system bandwidth is given by \( M \Delta f \), and \( NT_s \) corresponds to the OTFS frame duration. Also, $\Delta f$ is the subcarrier frequency and $T_c$ is the chip duration. The relationship between symbol duration and chip duration is expressed as $T_s = L T_c$. Here, $T_s$ is the symbol duration, and $L$ is the number of chips per symbol. Moreover, the DD domain channel response $h(\tau, \nu)$ is expressed as a sum of $P$ propagation paths, each defined by delay $\tau_u$, Doppler shift $\nu_u$, and coefficient $h_u$, i.e., $h(\tau, \nu) = \sum_{u=1}^{P} h_u \delta(\tau - \tau_u) \delta(\nu - \nu_u)$ \cite{OTFSILK}.
The parameters $\tau_u$ and $\nu_u$ are normalized as $\tau_u = \frac{k_u}{M\Delta f}, \quad \nu_u = \frac{l_i}{NT_s}.$ 

In the proposed system, the real and imaginary components of the information symbol vector $\textbf{s} = \textbf{s}_\Re+j\textbf{s}_\Im \in \mathbb{C}^{NM \times 1}$ in the DD domain are expressed as follows:
\begin{eqnarray}\label{transmission_signal_Re}
\textbf{s}_\Re = \Big[ s_\Re[1,1], s_\Re[1,2], s_\Re[1,3] \cdots s_\Re[N,M]
	\Big]^T,
\end{eqnarray}
\vspace{-1.5em}
\begin{eqnarray}\label{transmission_signal_Im}
\textbf{s}_\Im = \Big[ s_\Im[1,1], s_\Im[1,2], s_\Im[1,3] \cdots s_\Im[N,M]
	\Big]^T,
\end{eqnarray}
where the vectors $\textbf{s}_\Re$ and $\textbf{s}_\Im$ contain $NM$ complex $M_q$-QAM symbols. From the DD domain information symbols $s_\Re[k, l]$, $s_\Im[k, l]$, the real and imaginary components of the time-frequency domain information symbols, $S^\ell_{\Re}[n, m]$, $S^\ell_{\Im}[n, m]$, are obtained by inverse symplectic finite Fourier transform (ISFFT) as follows:
\begin{equation}
S^\ell_{\Re}[n, m] = \frac{1}{\sqrt{NM}} \sum_{k=0}^{N-1} \sum_{l=0}^{M-1} z_{c_{\Re},\ell} s_{\Re}[k, l] e^{j2\pi\left(\frac{nk}{N} - \frac{ml}{M}\right)}.
\end{equation}
\begin{equation}
S^\ell_{\Im}[n, m] = \frac{1}{\sqrt{NM}} \sum_{k=0}^{N-1} \sum_{l=0}^{M-1} z_{c_{\Im},\ell} s_{\Im}[k, l] e^{j2\pi\left(\frac{nk}{N} - \frac{ml}{M}\right)}.
\end{equation}
Then, by applying the Heisenberg transform to the information symbols in the time-frequency domain, the transmission-ready real and imaginary components of information symbols $x(t)$ in the time domain are expressed as follows:
\begin{equation}
x^\ell_{\Re}(t) = \sum_{n=0}^{N-1} \sum_{m=0}^{M-1} S^\ell_{\Re}[n, m] g_{\text{tx}}(t - nT) e^{j2\pi m\Delta f(t-nT)},
\end{equation}
\begin{equation}
x^\ell_{\Im}(t) = \sum_{n=0}^{N-1} \sum_{m=0}^{M-1} S^\ell_{\Im}[n, m] g_{\text{tx}}(t - nT) e^{j2\pi m\Delta f(t-nT)},
\end{equation}
where $g_{\text{tx}}$ is the transmit pulse.

As seen in the proposed system model, first, the $\eta_{OTFS-CIM}=NM\big(\log_2(M_q) + 2\log_2(N_C) \big)$ data bits in DD domain to be transmitted in a transmission period are divided into $NM$ DD grid subsets, each containing $\log_2(M_q)+2\log_2(N_C)$ bits. For each DD grid, out of $\log_2(M_q) + 2\log_2(N_C)$ bits, $n_q=\log_2(M_q)$ bits select the transmitted $M_q$-QAM symbol $s$, and $n_c = 2\log_2(N_C)$ bits select the active spreading code indices $c_\Re$ and $c_\Im$. The QAM symbol $s$ is divided into its real $s_\Re$ and imaginary $s_\Im$ components. Subsequently, the real part $s_\Re$ of the $s$ symbol is multiplied by the active spreading code $\textbf{z}_{c_\Re}$ corresponding to the $I$ component, and the imaginary part $s_\Im$ of the $s$ symbol is multiplied by the active spreading code $\textbf{z}_{c_\Im}$ corresponding to the $Q$ component. Finally, the signals $s_\Re \textbf{z}_{c_\Re}$ and $s_\Im \textbf{z}_{c_\Im}$ are transmitted by one transmit antenna at the transmitter through a Rayleigh fading channel to $N_R$ receive antennas. The mapping procedure of the proposed OTFS-CIM system is presented in Table \ref{OTFS_CIM_MAP} for different system parameters.  In Table \ref{OTFS_CIM_MAP}, ${[}00011001...1111{]}$ data bits are assumed to be transmitted in a transmission period with the proposed OTFS-CIM system and a numerical example is considered for the system parameters $M_q=4$, $N_{C}=2$, $N=4$, and $M=4$.

In the proposed OTFS-CIM system, information symbols are assumed to be transmitted through a time-varying multipath Rayleigh fading channel, where various channel delays and Doppler shifts characterize each path. The time-domain channel matrix of the proposed OTFS-CIM system denoted as $\textbf{G}(t) \in \mathbb{C}^{N_RNM \times NM}$, is expressed as follows:
\begin{equation}
    \textbf{G}(t)  = \begin{bmatrix}
        h_{1,1} & h_{1,2} & \cdots & h_{1,NM} \\
        \ h_{2,1} & h_{2,2} & \cdots & h_{2,NM} \\
        \ \vdots & \vdots & \vdots & \vdots \\
        \ h_{N_RNM,1} & h_{N_RNM,2} & \cdots & h_{N_RNM,NM}
    \end{bmatrix}.
\end{equation}



The effective channel matrix in the DD domain of the OTFS-CIM system is expressed as follows \cite{OTFS_SM1}:
\begin{equation}
\mathbf{H}^{\text{eff}} = \left( \mathbf{F}_N \otimes \mathbf{I}_{N_RM} \right) \mathbf{G}(t) \left( \mathbf{F}^{T}_N \otimes \mathbf{I}_{M} \right),
\end{equation}
where  $\mathbf{I}_{N_RM}$ is an identity matrix of size $N_RM$, while $\mathbf{I}_{M}$ is an identity matrix of size $M$. Also, $\mathbf{F}_N$ is the DFT matrix of size $N$.

For the $I$ and $Q$ components reaching the receiver of the proposed OTFS-CIM system, the $\ell^\text{th}$ chip signals $\textbf{y}^I_{\ell,v} \in \mathbb{C}^{N_RNM \times 1}$ and $\textbf{y}^Q_{\ell,v} \in \mathbb{C}^{N_RNM \times 1}$ for the $v^\text{th}$ DD grid can be expressed as follows:
\begin{eqnarray}\label{receive3}
	 \textbf{y}^I_{\ell,v} &   = & s_{\Re} z_{{c_\Re},\ell }\textbf{h}_{v}  + \mathbf{w}^{\text{eff}}_I,   \\
	  \textbf{y}^Q_{\ell,v}   &   = &  s_{\Im} z_{{c_\Im},\ell }\textbf{h}_{v}  + \mathbf{w}^{\text{eff}}_Q,   
\end{eqnarray}
where $\textbf{h}_{v} \in \mathbb{C}^{N_RNM \times 1}$ is the $v^\text{th}$ column of $\mathbf{H}^{\text{eff}}$. Also, $s_{\Re}$ and $s_{\Im}$ are the real and imaginary components of the complex QAM symbol $s$, respectively. $z_{c_{\Re},\ell}$ and $z_{c_{\Im},\ell}$ represent the $\ell^{\text{th}}$  chips of the spreading codes $\textbf{z}_{c_{\Re}}$ and $\textbf{z}_{c_{\Im}}$ , respectively. Also, the effective noise vectors, denoted as $\mathbf{w}^{\text{eff}}_I$ and $\mathbf{w}^{\text{eff}}_Q$, are expressed as follows:
\begin{equation}
\mathbf{w}^{\text{eff}}_I = \left( \mathbf{F}_n \otimes \mathbf{I}_{N_RM} \right) \mathbf{w}_I,
\end{equation}
\begin{equation}
\mathbf{w}^{\text{eff}}_Q = \left( \mathbf{F}_n \otimes \mathbf{I}_{N_RM} \right) \mathbf{w}_Q,
\end{equation}
where $\mathbf{w}_I$ and $\mathbf{w}_Q$ are complex additive white Gaussian noise (AWGN) vectors with zero mean and variance \(N_0\) for the \(I\) and \(Q\) components, respectively.

A despreading-based ML detector is used to estimate the transmitted symbol ($s$) and the active spreading code indices ($c_{\Re}$ and $c_{\Im}$) from the signals received by the receiver terminal of the proposed CIM-OTFS system. Unlike the ML detector, which considers all possible values of the parameters to be estimated simultaneously, the despreading-based ML detector uses a sequential approach. It starts by estimating the spreading code indices and then estimates the transmitted symbol using the previously estimated spreading code indices. This approach significantly reduces the overall complexity by ensuring that the computational burden associated with spreading code estimation is added to the overall complexity additively rather than multiplicatively. At the receiver terminal of the proposed CIM-OTFS system, the received signal is initially decomposed into $I$ and $Q$ components, as shown in Fig. \ref{system_model}. The indices of the spreading codes that maximize the $I$ and $Q$ components of the received signal are selected as the estimated spreading code indices $\hat{c}_\Re$ and $\hat{c}_\Im$, respectively. Then, the ML detector estimates the transmitted QAM symbol using the maximized received signals corresponding to these estimated spreading code indices.

The values $\tilde{\textbf{y}}^I_{v} \in \mathbb{C}^{N_RNM \times 1}$ and $\tilde{\textbf{y}}^Q_{v} \in \mathbb{C}^{N_RNM \times 1}$ at the $c^\text{th}$ correlator output for the $I$ and $Q$ components, respectively, are defined as follows:
\begin{equation}\label{corelator1}
\tilde{\textbf{y}}^I_{v,c}     =  \textbf{Y}^I_{v} \textbf{z}_{c}, \, \, \,  \tilde{\textbf{y}}^Q_{v,c}     =  \textbf{Y}^Q_{v} \textbf{z}_{c},
\end{equation}
where $\textbf{Y}^I_{v} \in \mathbb{C}^{N_RNM \times L}=\Big[\textbf{y}^I_{1,v}\, \textbf{y}^I_{2,v}\, \ldots \,\textbf{y}^I_{L,v}\Big]$ and $\textbf{Y}^Q_{v} \in \mathbb{C}^{N_RNM \times L}=\Big[\textbf{y}^Q_{1,v}\, \textbf{y}^Q_{2,v}\, \ldots \,\textbf{y}^Q_{L,v}\Big]$. The $\tilde{y}^{I,r}_{v,c}  \in \mathbb{C}^{1 \times 1}$ and $\tilde{y}^{Q,r}_{v,c}  \in \mathbb{C}^{1 \times 1}$ expressions for the $r^\text{th}$ receive antenna are denoted as follows: 
\begin{eqnarray}\label{eq8} 
\tilde{y}^{I,r}_{v,c} &  = & \sum_{\ell=1}^{L} z_{c,\ell}\,y^{I,r}_{v,\ell} = \sum_{\ell=1}^{L} z_{c,\ell}\,\Big(s_{\Re}\,z_{c_\Re,\ell}\, h_{v}^{r}  + w^{\text{eff}}_{I,r,c} \Big), \nonumber \\
&=&
\begin{cases}s_{\Re}\, h_{v}^{r} \,E_z  + \bar{w}^{\text{eff}}_{I,r,c}, & \text{if} \:\:\: c =c_\Re \\
\bar{w}^{\text{eff}}_{I,r,c}, &  \text{if} \:\:\: c \not=c_\Re \end{cases} \\   \label{eq7_2}
\tilde{y}^{Q,r}_{v,c}  &  = & \sum_{\ell=1}^{L} z_{c,\ell}\,y^{Q,r}_{v,\ell} = \sum_{\ell=1}^{L} z_{c,\ell}\,\Big(s_{\Im}\,z_{c_\Im,\ell}\, h_{v}^{r}  + w^{\text{eff}}_{Q,r,c}\Big), \nonumber \\
&=&
\begin{cases}s_{\Im}\, h_{v}^{r}\, E_z  + \bar{w}^{\text{eff}}_{Q,r,c}, & \text{if} \:\:\: c =c_\Im \\
\bar{w}^{\text{eff}}_{Q,r,c}, &  \text{if} \:\:\: c \not=c_\Im \end{cases} 
\end{eqnarray}
where $r \in \{1, 2, 3 \dots, N_R\}$ and $c \in \{1, 2, 3 \dots, N_C\}$. The average energy transmitted per symbol is represented by $E_z$, which is defined as: $E_z=\frac{1}{L}\sum_{\ell=1}^{L}{z^2_{c,\ell}}$. Furthermore, $\bar{w}^{\text{eff}}_{I,r,c}$ and $\bar{w}^{\text{eff}}_{Q,r,c}$ are given as:
\begin{equation}
    \bar{w}^{\text{eff}}_{I,r,c} = \frac{1}{L}\sum_{\ell=1}^{L} z_{c,\ell} w^{\text{eff}}_{I,r,c}, \ \  \bar{w}^{\text{eff}}_{Q,r,c} = \frac{1}{L}\sum_{\ell=1}^{L} z_{c,\ell} w^{\text{eff}}_{Q,r,c}.
\end{equation}
To determine the spreading code indices $c_\Re$ and $c_\Im$ for the $I$ and $Q$ components, as shown in Fig. \ref{system_model}, a method is applied that involves computing the squared norm of the received signal vector. The indices corresponding to the maximum squared norm are selected as the estimated spreading code indices $\hat{c}_\Re$ and $\hat{c}_\Im$. This estimation of spreading codes is performed using the orthogonality of the WH codes. The orthogonality of WH codes is defined as $\sum_{\ell=1}^{L}z_{i_1,\ell}\, z_{i_2,\ell}   = \bigg\{\begin{array}{cc}
1, & \text{if}\,\,\,\,i_1 = i_2\\ 
0, & \text{if}\,\,\,\,i_1 \neq i_2 
\end{array}$ .
As a result, the active spreading code indices $c_\Re$ and $c_\Im$ are estimated as follows:
\begin{equation}\label{eq10}
\hat{c}_\Re    =  \underset{c}{\mathrm{arg\,max}} \ \big|\big| \tilde{\textbf{y}}^{I}_{v,c}\big|\big|^2 , \ \ \hat{c}_\Im      =    \underset{c}{\mathrm{arg\,max}} \ \big|\big| \tilde{\textbf{y}}^{Q}_{v,c} \big|\big|^2. 
\end{equation}
The estimated spreading code indices $\hat{c}_\Re$, $\hat{c}_\Im$ are utilized to obtain the values of $\tilde{\textbf{y}}^{I}_{v,\hat{c}_\Re}$ and $\tilde{\textbf{y}}^{Q}_{v,\hat{c}_\Im}$ by substituting them into the received signal vector. The ML detector of the proposed OTFS-CIM system, which estimates the QAM symbol $s$ using the estimated spreading code indices $\hat{c}_\Re$ and $\hat{c}_\Im$, is defined as follows: 
\begin{eqnarray}\label{eq11}
 \big[ \hat{s}_\Re, \hat{s}_\Im  \big] \!\! = \! \text{arg}\underset{\hat{s}_\Re, \hat{s}_\Im}{\mathrm{min}} \ \!\! \Big|\Big| \! \Big(\tilde{\textbf{y}}^{I}_{v,\hat{c}_\Re}\!\!+\!j\tilde{\textbf{y}}^{Q}_{v,\hat{c}_\Im}\Big) \! - \! E_z (\hat{s}_\Re\!+\!j\hat{s}_\Im) \textbf{h}_{v}\Big|\Big|^2  \!\!\!.\!
\end{eqnarray}
In OTFS-CIM, the ML detector estimates the transmitted QAM symbol $s$ for each point in the DD grid. This process is repeated $N M$ times to estimate $s$ for the entire DD grid. Then, the $NM$ estimated QAM symbol is obtained as follows:
\begin{eqnarray}\label{eq15}
\hat{\textbf{s}_\Re} = \Big[\hat{s}_\Re[1,1], \hat{s}_\Re[1,2], \cdots, \hat{s}_\Re[N,M]\Big]^T.
\end{eqnarray}
\begin{eqnarray}\label{eq16}
\hat{\textbf{s}_\Im} = \Big[\hat{s}_\Im[1,1], \hat{s}_\Im[1,2], \cdots, \hat{s}_\Im[N,M]\Big]^T.
\end{eqnarray}
As a result, based on the values of $\hat{s}_\Re$,$\hat{s}_\Im$, $\hat{c}_\Re$ and $\hat{c}_\Im$, the receiver of the OTFS-CIM system uses the bit demapper block shown in Fig. \ref{system_model} to generate the estimated data bits in the DD domain.

\begin{figure}[t]
\centering{\includegraphics[width=0.4\textwidth]{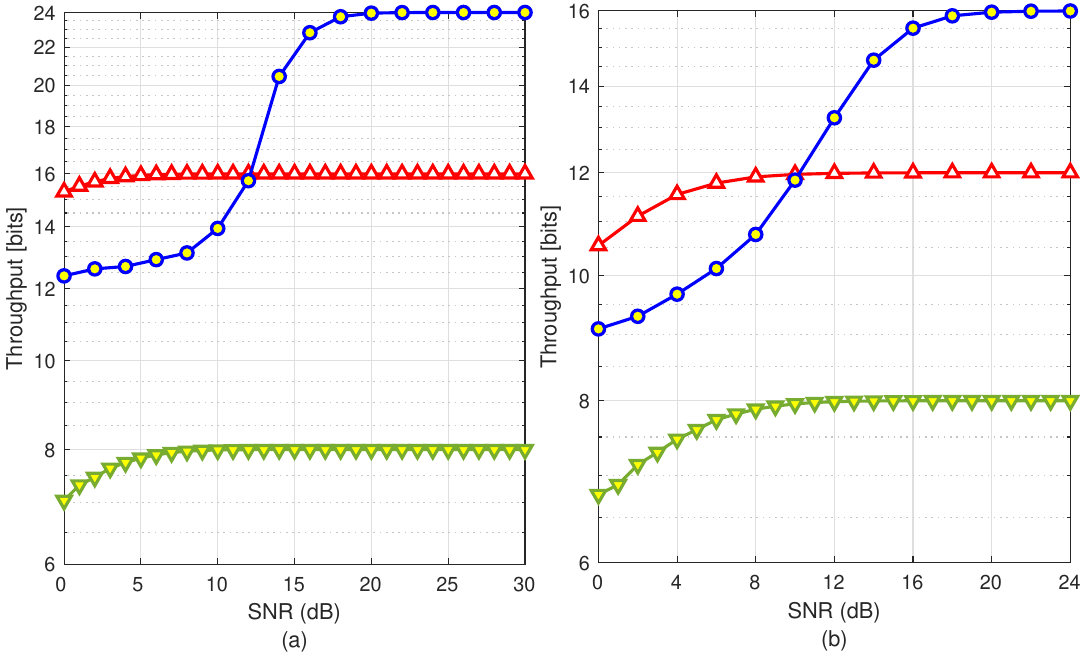}}\vspace{-1em}
	\caption{Throughput comparisons of the proposed OTFS-CIM, OTFS-SM, and OTFS systems.} 
	\label{thro} 
     \vspace{-1.5em}
\end{figure}

\section{Performance analyses}

The performance analyses presented in this section aim to comprehensively evaluate the proposed OTFS-CIM system by examining its throughput, spectral efficiency, and energy-saving analysis relative to benchmark systems, including OTFS and OTFS-SM.

\subsection{Throughput Analysis}
Throughput is a key performance metric in wireless communication systems that measures the successful data transmission rate over a wireless channel. It is defined as the number of data bits out of the total bits transmitted that can only be obtained correctly at the receiving end, considering errors or losses during transmission. The throughput of the proposed OTFS-CIM system can be defined as $\mathcal{\zeta} = \frac{\big(1 - \text{BER}_{\text{OTFS-CIM}}\big)}{T_s} \eta_{\text{OTFS-CIM}}$ \cite{Tse}, 
where \(\big(1 - \text{BER}_{\text{OTFS-CIM}}\big)\) represents the probability of correctly decoding the transmitted bits during the symbol transmission duration \(T_s\) in the OTFS-CIM system. Throughput analysis shows that the proposed OTFS-CIM system achieves significantly higher throughput than benchmark systems such as OTFS and OTFS-SM. This is supported by the results in Fig. \ref{thro}, which demonstrate the superior throughput of the proposed OTFS-CIM system. Fig. \ref{thro} illustrates the throughput comparisons among OTFS, OTFS-SM, and OTFS-CIM systems. In Fig. \ref{thro} (a), the system parameters $M_q=4$, $N_T=4$, $N_C=4$, $L=8$, $N=2$, $M=2$, and $N_R=3$ are utilized, while Fig. \ref{thro} (b) employs the system parameters $M_q=4$, $N_T=2$, $N_C=2$, $L=4$, $N=2$, $M=2$, and $N_R=2$. The results demonstrate that the proposed OTFS-CIM system achieves higher throughput than the OTFS and OTFS-SM systems.

\begin{figure}[t]
\centering{\includegraphics[width=0.35\textwidth]{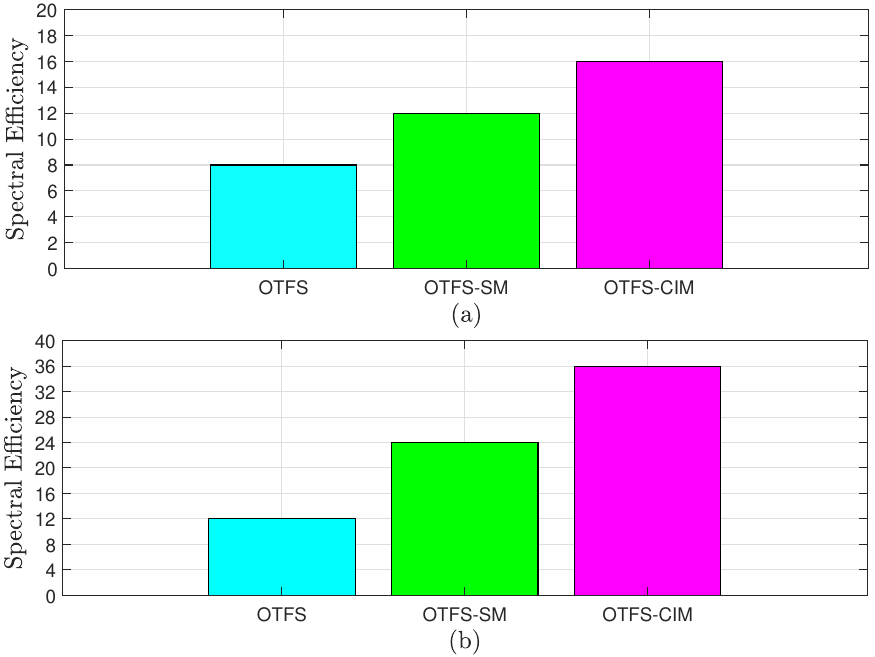}}\vspace{-1em}
	\caption{Spectral efficiency comparisons of the proposed OTFS-CIM, OTFS-SM, and OTFS systems for (a) $M_q=4$, $N_T=2$, $N_C=2$ and (b) $M_q=8$, $N_T=8$, $N_C=8$.} 
     \vspace{-1em}
	\label{spec} 
\end{figure}

\subsection{Data Rate Analysis}

Spectral efficiency describes the maximum data rate achievable over a bandwidth. A comparative analysis of spectral efficiencies among various wireless communication systems, including the proposed OTFS-CIM, OTFS-SM, and OTFS, is illustrated in Fig. \ref{spec}. In Fig. \ref{spec} (a), the system parameters $M_q=4$, $N_T=2$, $N_C=2$, $N=2$, $M=2$ are utilized, whereas in Fig. \ref{spec} (b), the system parameters $M_q=8$, $N_T=8$, $N_C=8$, $N=2$, $M=2$ are employed. The spectral efficiencies of OTFS and OTFS-SM systems are expressed as $\eta_\text{{OTFS}}=NM\big(\log_2(M_q) \big)$ and $\eta_\text{{OTFS-SM}}=NM\big(\log_2(M_q N_T)\big)$, respectively. The spectral efficiency of the proposed OTFS-CIM system is already given in (\ref{spec_cim}). Fig. \ref{spec} indicates that the proposed OTFS-CIM system demonstrates the most dramatic enhancement in spectral efficiency compared to the other systems, particularly when system parameter values are increased. Therefore, it is concluded that the proposed OTFS-CIM system can transmit data at a higher data rate than OTFS and OTFS-SM systems.

\begin{figure}[t]
\centering{\includegraphics[scale=0.4]{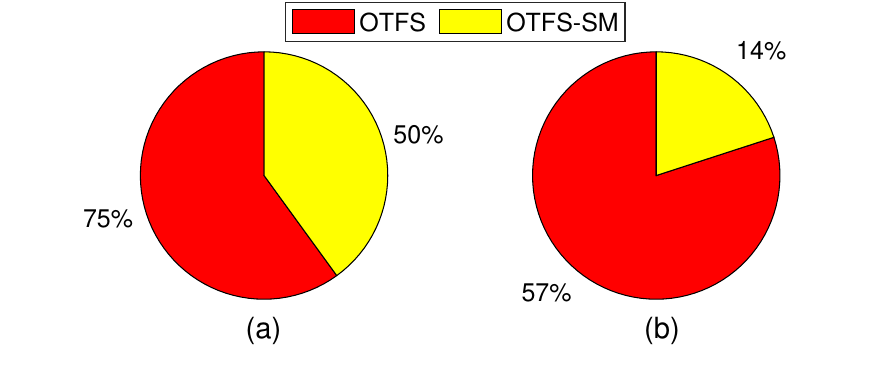}}\vspace{-1em}
	\caption{Energy saving percentages of the OTFS-CIM system compared to OTFS and OTFS-SM systems for (a) $M_q=4$, $N_T=4$, $N_C=8$, $N=2$, $M=2$ and (b) $M_q=8$, $N_T=8$, $N_C=4$, $N=4$, $M=4$.}
	\label{energ} 
    \vspace{-1.5em}
\end{figure}

\subsection{Energy Saving Analysis}
Wireless communication systems are increasingly required to improve their energy efficiency due to the escalating energy consumption costs and the need to reduce carbon emissions for environmental sustainability. In this context, the proposed OTFS-CIM system presents an efficient solution that uses index-based data transmission to achieve energy savings. The energy saving percentage $\mathbb{E}_{\text{sav}}$ per $\eta_\text{OTFS-CIM}$ bits of the proposed OTFS-CIM system is described as $\mathbb{E}_{\text{sav}} = \Big(1-\frac{\eta_{\text{b}}}{\eta_{\text{OTFS-CIM}}}\Big)E_b\%$, 
where $E_b$ represents the energy per bit and $\eta_{\text{b}}$ denotes the spectral efficiency of the benchmark systems (OTFS and OTFS-SM). The energy efficiency gains of the proposed OTFS-CIM system are illustrated in Fig. \ref{energ} for different system parameters. It is shown that the OTFS-CIM system has higher energy efficiency than OTFS and OTFS-SM for different system parameters. In Fig. \ref{energ} (a), the system parameters $M_q=4$, $N_T=4$, $N_C=8$, $N=2$, and $M=2$ are used, while in Fig. \ref{energ} (b), the system parameters $M_q=8$, $N_T=8$, $N_C=4$, $N=4$, and $M=4$ are used. For example, in Fig. \ref{energ} (a), the OTFS-CIM system has energy savings of $75\%$ and $50\%$ compared to OTFS and OTFS-SM systems.

\begin{figure}[t]
\centering{\includegraphics[scale=0.44]{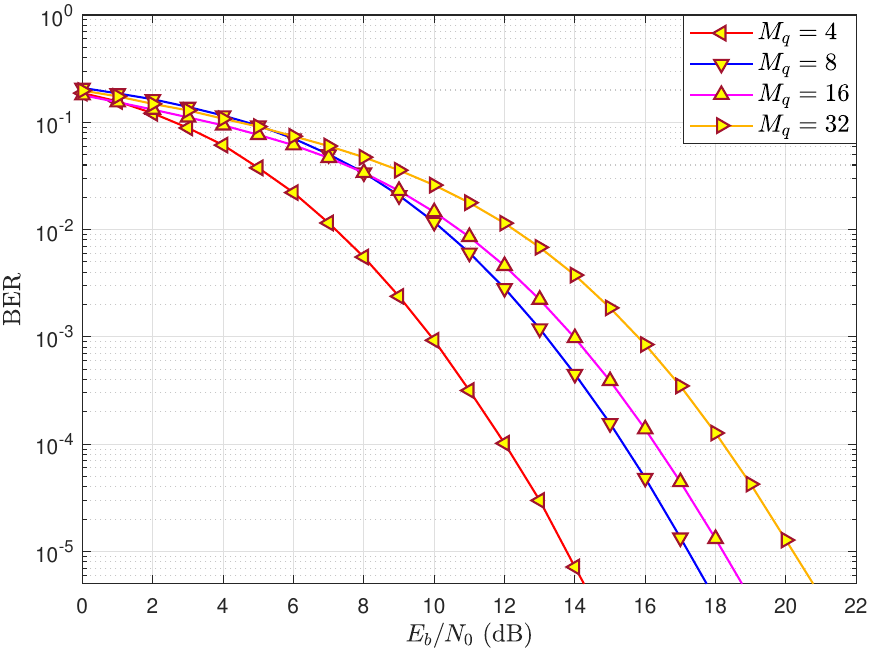}}	\vspace{-1em}
	\caption{Error performance curves of the proposed OTFS-CIM for different $M_q$ values.}
	\label{OTFS_CIM_Mqs} 
     \vspace{-1.5em}
\end{figure}

\begin{figure}[t]
\centering{\includegraphics[scale=0.44]{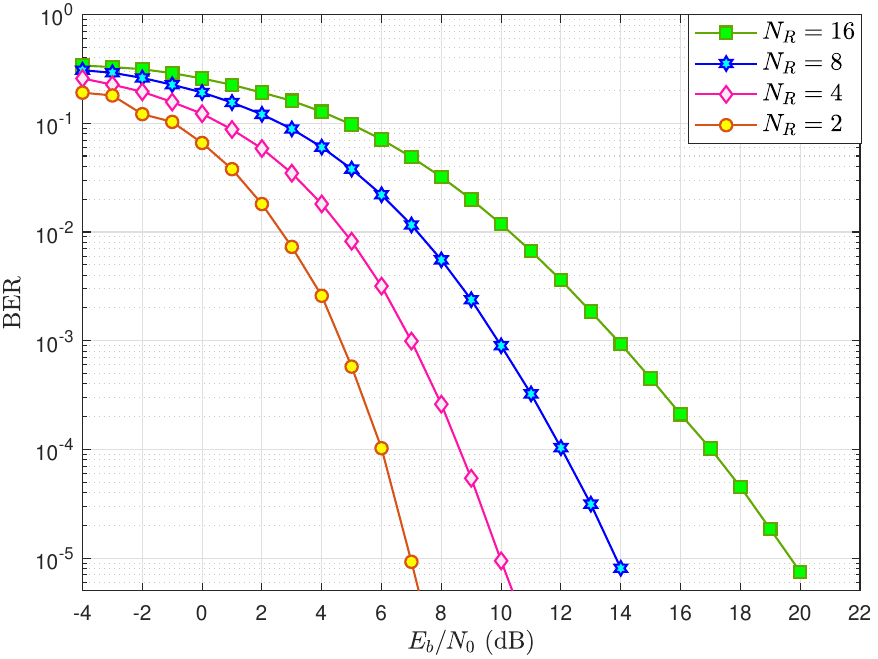}}\vspace{-1em}	
	\caption{Error performance curves of the proposed OTFS-CIM for different $N_R$ values.}
	\label{OTFS_CIM_NRs} 
     \vspace{-1.5em}
\end{figure}

\section{Simulation Results}
This section presents simulation results of the OTFS-CIM system obtained by combining OTFS and CIM techniques in a SIMO architecture for high-mobility Rayleigh fading channels. The comparative error performance curves of the proposed OTFS-CIM, OTFS-SM, and OTFS systems using $M_q$-QAM modulation are given for different system parameters. It is seen that the proposed system provides better error performance than OTFS and OTFS-SM systems for all parameters and conditions considered. In all simulations, the SNR is represented as $\mathrm{SNR (dB)} = 10\log_{10}(E_s/N_0)$. Also, the system parameters are as follows: the carrier frequency is $4$ GHz, the subcarrier spacing is $15$ KHz, the maximum speed is $506.2$ km/h, and the channel tap count is $4$.

The effect of the value of the modulation order, represented as $M_q$, on the error performance of the OTFS-CIM system is presented in Fig. \ref{OTFS_CIM_Mqs}. The system parameters are selected as $M_q=(4,8,16,32)$, $N_C=2$, $L=8$, $N_R=4$, $N=2$, and $M=2$, respectively. As shown in Fig. \ref{OTFS_CIM_Mqs}, the error performance of the OTFS-CIM system deteriorates with increasing $M_q$. Furthermore, increasing $M_q$ enhances the spectral efficiency of the OTFS-CIM system by $NM \log_2(M_q)$. Therefore, considering the data rate and error performance requirements, a reasonable $M_q$ value should be determined.

\begin{figure}[t]
\centering{\includegraphics[scale=0.44]{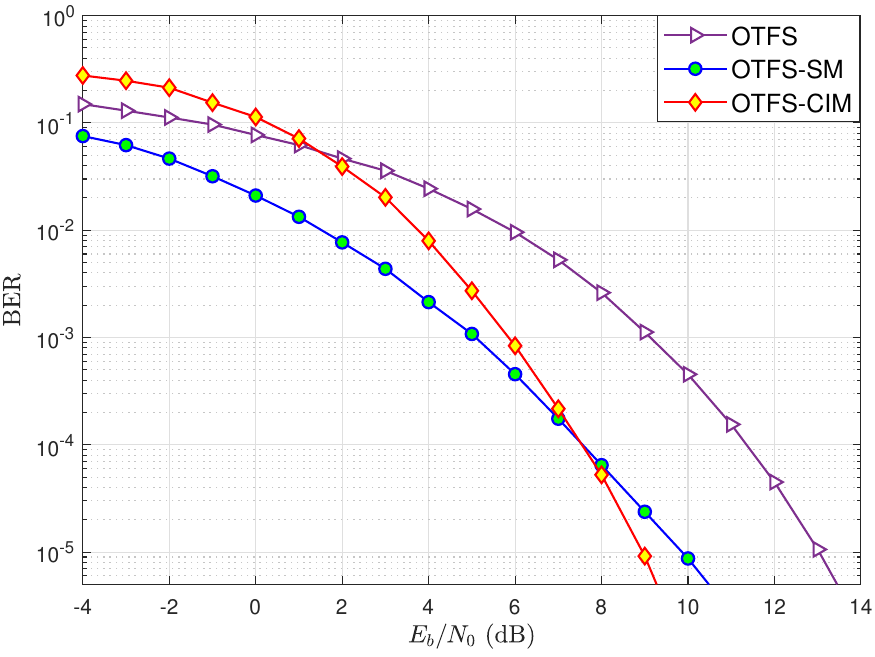}}	\vspace{-1em}
	\caption{Error performance comparison of the proposed OTFS-CIM, OTFS-SM, and OTFS systems for $\eta_\text{{OTFS-CIM}} = 24$ bpcu.}
	\label{OTFS_CIM_24} 
     \vspace{-1em}
\end{figure}

\begin{figure}[t]
\centering{\includegraphics[scale=0.44]{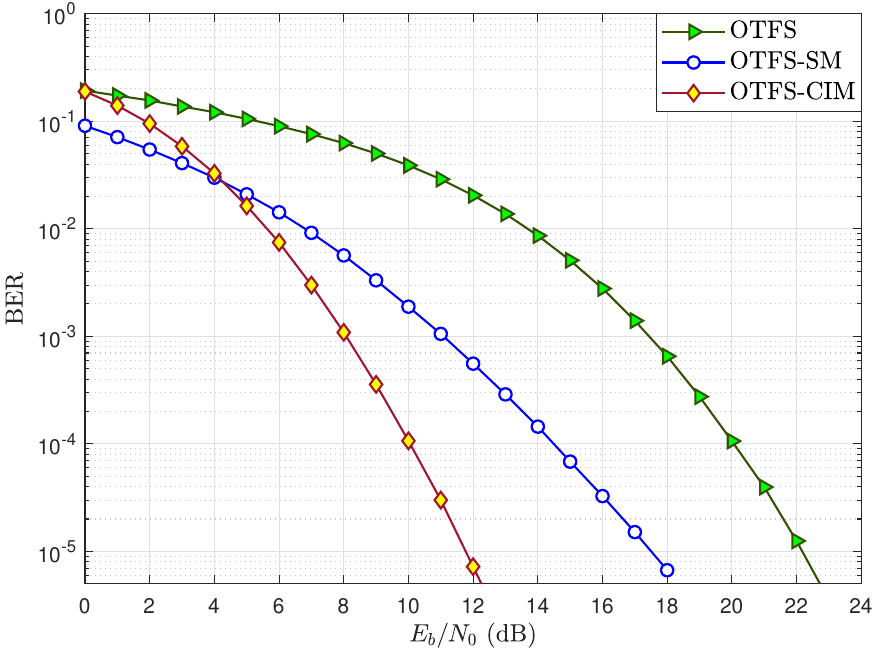}}	\vspace{-1em}
	\caption{Error performance comparison of the proposed OTFS-CIM, OTFS-SM, and OTFS systems for $\eta_\text{{OTFS-CIM}} = 32$ bpcu.}
     \vspace{-1.5em}
	\label{OTFS_CIM_32} 
\end{figure}

The BER performance of the proposed OTFS-CIM system is shown in Fig. \ref{OTFS_CIM_NRs} for varying numbers of receive antennas ($N_R$). Fig. \ref{OTFS_CIM_NRs} demonstrates that increasing $N_R$ significantly improves error performance. However, increasing the number of $N_R$ enhances error performance at the expense of greater receiver complexity, creating a trade-off between performance and complexity. The spatial diversity afforded by multiple receive antennas effectively mitigates multipath fading and inter-Doppler interference inherent in high-mobility channels.

Fig. \ref{OTFS_CIM_24} presents the error performance of the proposed OTFS-CIM system compared to conventional OTFS and OTFS-SM schemes over the Rayleigh channel with Doppler spread for spectral efficiency of $24$ bpcu. The proposed OTFS-CIM system employs $M_q=4$, $N_C=4$, and $L=8$  system parameters, whereas the OTFS-SM and OTFS systems utilize $M_q=16$, $N_T=4$, and $M_q=64$ system parameters, respectively. Also, $N=2$, $M=2$, and $N_R=8$ system parameters are selected for all systems. As seen in Fig. \ref{OTFS_CIM_24}, the proposed OTFS-CIM system demonstrates better error performance in high-mobility scenarios with Doppler spreads compared to OTFS and OTFS-SM systems.

Fig. \ref{OTFS_CIM_32} shows the error performance comparison of the OTFS-CIM system against OTFS and OTFS-SM schemes for spectral efficiency of $32$ bpcu. While the OTFS-CIM system employs system parameters ($M_q=4$, $N_C=8$, and $L=16$) to improve spectral and energy efficiency, the OTFS-SM and OTFS systems utilize system parameters ($M_q=32$, $N_T=8$) and $M_q=256$, respectively. The system parameters $N=2$, $M=2$, and $N_R=4$ are chosen for all systems. In Fig. \ref{OTFS_CIM_32}, BER curves demonstrate that the OTFS-CIM outperforms OTFS and OTFS-SM systems. These results show the robustness of the OTFS-CIM system in high-mobility communication scenarios, where severe Doppler shifts degrade system reliability.

\section{Conclusion}
The proposed OTFS-CIM system, which combines OTFS modulation with the CIM technique in a SIMO architecture, has been comprehensively evaluated on high-mobility Rayleigh channels under various system configurations. It has been observed that increasing the modulation order ($M_q$) has negatively affected system performance, while adding more receive antennas has significantly enhanced it. The analysis results have indicated that the proposed system has outperformed conventional OTFS and OTFS-SM systems in throughput, spectral efficiency, and energy saving. Furthermore, the proposed system has demonstrated improved error performance in all tested scenarios, highlighting its potential as a solution for next-generation wireless communication in high-mobility environments.

\vspace{-0.5em}



\bibliographystyle{IEEEtran}


\bibliography{Referanslar}

\end{document}